\documentclass[%
reprint,
nofootinbib,
amsmath,
amssymb,
prd,
longbibliography,
superscriptaddress,twocolumn,superscriptaddress,floats,floatfix,showpacs]{revtex4-2}

\usepackage{mathrsfs}
\usepackage{graphicx}
\usepackage{dcolumn}
\usepackage{bm}
\usepackage{lipsum}
\usepackage{graphics}
\usepackage{bm}
\usepackage{bbm}
\usepackage{color}
\usepackage{microtype}
\usepackage{IEEEtrantools}
\usepackage{mathrsfs}
\usepackage{amsthm}
\usepackage{enumitem}
\usepackage[normalem]{ulem}
\usepackage{graphics}
\usepackage[T1]{fontenc}

\usepackage[dvipsnames, usenames]{xcolor}
\usepackage[unicode, colorlinks=true, linkcolor=linkcolor, citecolor=linkcolor, filecolor=linkcolor,urlcolor=linkcolor, pdfusetitle]{hyperref}
\usepackage[all]{hypcap}
\usepackage[utf8]{inputenc}
\usepackage{orcidlink}

\definecolor{linkcolor}{rgb}{0.0,0.3,0.5}
\definecolor{dodgerblue}{HTML}{1E90FF}
\definecolor{darkgreen}{rgb}{0,0.5,0}
\definecolor{romared}{RGB}{142,0,28}

\usepackage{hyperref}
\hypersetup{
    pdfnewwindow=true,      
    colorlinks=true,       
    linkcolor=romared,          
    citecolor=romared,        
    filecolor=romared,      
    urlcolor=romared        
}

\newcommand{\GSSI}{Gran Sasso Science Institute (GSSI), I-67100 L’Aquila, Italy}
\newcommand{\GranSasso}{INFN, Laboratori Nazionali del Gran Sasso, I-67100 Assergi, Italy}

\renewcommand{\O}{\mathcal{O}}

\newcommand{\order}[1]{\mathcal{O}(#1)}

\def\m{\mathtt{m}}

\renewcommand{\S}{{\mathcal S}}

\usepackage[T1]{fontenc}

\def\g{\mathbf{g}}
\def\G{\mathcal{G}}

\def\grav{\mathrm{grav}}
\def\p{\mathrm{p}}
\def\c{\mathrm{c}}
\def\m{\mathrm{m}}
\def\scal{\mathrm{scal}}
\def\diss{\mathrm{diss}}
\def\cons{\mathrm{cons}}

\def\R{\mathcal{R}}
\def\S{\mathcal{S}}

\begin{document}

\title{Measuring scalar charge with compact binaries: High accuracy modelling with self-force}

\author{Andrew Spiers}
\affiliation{Nottingham Centre of Gravity \& School of Mathematical Sciences, University of Nottingham, University Park, Nottingham, NG7 2RD, UK}
\author{Andrea Maselli}
\affiliation{\GSSI}
\affiliation{\GranSasso}
\author{Thomas P. Sotiriou}
\affiliation{Nottingham Centre of Gravity \& School of Mathematical Sciences, University of Nottingham, University Park, Nottingham, NG7 2RD, UK}
\affiliation{School of Physics and Astronomy, University of Nottingham, University Park, Nottingham, NG7 2RD, UK}

\date{\today}

\begin{abstract}
Using the self-force approach, we present the premier first-post-adiabatic accuracy formalism for modelling compact binaries in theories with a massless scalar field non-minimally coupled to gravity. We limit the binary secondary to being a non-spinning compact body with no scalar dipole (we will address the spinning and scalar dipole cases in an upcoming paper). By producing an ansatz for the scalar charged point particle action, we derive first- and second-order perturbative field equations and equations of motion for the secondary compact object. Under our assumptions, implementing this formalism will produce sufficiently accurate waveform templates for precision measurements of the scalar charge of the secondary with LISA data on extreme-mass-ratio inspirals. Our formalism is consistent with almost general scalar-tensor theories of gravity. Implementing our formalism builds on self-force models in General Relativity; we show the incorporation into the two-timescale formalism is straightforward. Excitingly, implementation poses no significantly more challenging barriers than computing first-post adiabatic waveforms in General Relativity.
\end{abstract}

\maketitle


\section{\label{sec:level1} Introduction}

The detection of a new fundamental field via its imprint on compact 
objects and the gravitational wave signals they produce would be a 
ground-breaking discovery. Indeed, such searches are among the key 
goals of current and future gravitational wave detectors \cite{Barausse:2020rsu,Kalogera:2021bya,LISA:2022kgy,Branchesi:2023mws,LIGOScientific:2021sio}. Asymmetric binaries are particularly promising sources in this context 
\cite{Yunes:2011aa,Barausse:2016eii,Datta:2019epe,Annulli:2020ilw,
Destounis:2020kss,Maggio:2021uge,Sago:2021iku,Collodel:2021jwi,Liang:2022gdk,Zhang:2023vok,Fell:2023mtf,Brito:2023pyl}. 
They consist of a larger body, the primary, of mass $M$, 
and a much smaller body, the secondary, of mass $\mu$\footnote{In this paper $\mu$ is the zeroth-order in $\varepsilon$ mass, we allow for higher order corrections to the mass which come from the presence of the scalar field.}. 
Their mass ratio
\begin{align}\label{eq:mass-ratio}
    \varepsilon=\frac{\mu}{M}\ ,
\end{align}
is then very small, with the more extreme cases, known as extreme 
mass ratio inspirals (EMRIs), reaching $\varepsilon\sim 10^{-6}$. 
EMRIs could execute some $10^5$ orbits while in LISA's band and 
be continuously observed for very long periods --- several months 
or even years. Hence, they are expected to be excellent probes 
of the properties of their source~\cite{amaro2015research}.

Until recently attention was strongly focused on EMRIs potential 
to probe the spacetime around, and hence the properties of, the 
primary 
 black hole \cite{Ryan:1995wh,Barack:2006pq,Glampedakis:2005cf}. However, 
it was recently pointed out in \cite{maselli2020detecting} that 
the scalar charge of the secondary could leave a strong imprint on the EMRI waveforms --- strong enough to make the charge measurable by LISA \cite{Maselli:2021men}. It has also been shown 
in \cite{maselli2020detecting}, using an effective field theory 
framework, that, to leading order in $\varepsilon$, the charge of 
the primary is negligible. Hence, the primary is adequately 
approximated by a Kerr black hole, and the charge for the secondary fully controls the deviations from GR. This drastically 
simpler framework has already been extended to eccentric orbits \cite{Barsanti:2022ana} and massive scalar fields \cite{Barsanti:2022vvl}. 
We will discuss the assumptions underpinning this framework in detail below, but its most important practical limitation is that it does not include any post-adiabatic corrections that appear at 
second order in $\varepsilon$.


Indeed, a burgeoning method for modelling compact binaries is the self-force approach in black hole perturbation theory. This method provides an accurate approximation in the extreme-mass-ratio limit, $\varepsilon\lesssim 10^{-4}$,
where it is exorbitantly computationally expensive to produce full inspiral waveforms using Numerical Relativity due to the binary's disparate length scales. Tackling this problem using perturbation theory is advantageous because $\varepsilon$ is a naturally small expansion parameter. World-leading perturbative self-force models are to first-post-adiabatic accuracy. Such accuracy provides waveforms with $\O(\varepsilon)$ phase error over the course of an inspiral of $\O(\frac{1}{\varepsilon})$ orbits. Currently, waveforms to first-post adiabatic accuracy have been computed for quasi-circular inspirals of Schwarzschild black holes in GR. These waveforms show impressive agreement with Numerical Relativity waveforms for mass-ratios smaller than $\O(\frac{1}{10})$. Significant efforts are ongoing to extend these results to generic orbits around a Kerr primary black hole~\cite{mySecond-orderTeuk, green2020teukolsky, toomani2021new, dolan2022gravitational, Leather-unpublished, Upton, Upton2, osburn2022new}, and include effects such as resonances~\cite{nasipak2021resonant} and the spin of the secondary~\cite{Piovano:2020zin,mathews2022self, drummond2022precisely1, drummond2022precisely2, drummond2023extreme}.

Extending the perturbative self-force approach to theoretical scenarios that include new fundamental fields is a fledgling field of research of clear importance. Constraining the existence of new fundamental fields with LISA measurements on EMRIs
requires high-accuracy waveform templates, including the effects of these new fundamental fields. As in GR, one has to reach first-post adiabatic accuracy.
As a first, critical step in this direction, in this paper, we push the approach of~\cite{maselli2020detecting, Maselli:2021men} to post-adiabatic order. By generalising the perturbative self-force approach and using a new ansatz for the point-particle action, we derive the field equations and equations of motion for the secondary. That is, we provide the required equations for modelling inspirals to first-post adiabatic accuracy in scenarios with a massless scalar field nonminimally coupled to gravity. This is the first scheme for modelling binaries perturbatively to first-post adiabatic accuracy beyond GR and the Standard Model and provides a roadmap to full calculations. 


The paper is organised as follows. In Sec.~\ref{sec:pointcharge}, we discuss the theoretical ground in which we model 
asymmetric binaries beyond GR. 
In Sec.~\ref{sec:SSF}, we derive a formalism for 
calculating the first- and second-order self-force in a large class of scalar-tensor theories of gravity.
We exploit our formalism to build a two-timescale approximation for efficient first-post adiabatic accurate modelling in Sec.~\ref{sec:two-timescale}. 
In appendix~\ref{app:SFGR}, we briefly review the self-force approach within black hole perturbation theory in GR, derive the metric perturbation field equations and perturbative equations of motion from our action approach, and demonstrate why the first-post adiabatic models provide high-accuracy waveforms ready for LISA observations.


\section{Asymmetric binaries and scalar fields}\label{sec:pointcharge}
%

\subsection{Action and field equations}

Following Ref.~\cite{maselli2020detecting}, our starting point will be a general action 
which describes a real scalar field $\varphi$ non-minimally coupled to gravity:
\begin{align}\label{eq:action}
    S[\g_{ab} , \varphi,\Psi] = S_0[\g_{ab} , \varphi] + \alpha S_\c[\g_{ab} , \varphi] + S_\m[\g_{ab} , \varphi,\Psi]\ ,
\end{align}
where 
\begin{align}\label{eq:S0}
    S_0[\g_{ab} , \varphi]=\int \frac{\sqrt{-\g}}{16\pi}\Big( R-\frac{1}{2}\partial_\mu\varphi\partial^\mu \varphi  \Big) d^4x\ ,
\end{align}
$\g$ is the metric determinant, and
we work in geometric units $G=c=\hbar=1$.
$S_\c$ contains any additional (self)-interactions for the scalar, whereas $S_\m$ describes the matter 
fields $\Psi$. For concreteness, we consider here a massless scalar, omit the bare mass term in $S_0$, and consistently assume that $S_\c$ respects shift symmetry. However, this approach can be straightforwardly generalized to massive scalars \cite{Barsanti:2022vvl}, and we will return to this point later. 

We take the coupling constant  
$\alpha$ to have dimensions $[{\rm length}]^n$ where $n\geq 2$. This corresponds to negative mass dimensions in particle physics units; {\em i.e.},~we assume that these interactions are suppressed by some characteristic energy scale. Note that there are no terms that would correspond to $n\leq 1$ that are consistent with local Lorentz symmetry \cite{Saravani:2019xwx}. We apply no other conditions to $S_\c$, and hence, our approach covers a very broad range of theoretical scenarios. 
%
%
%
For the time being, we leave $S_m$ generic.

To derive field equations for the metric, one varies the 
action \eqref{eq:action} with respect to the metric, 
that yields
\begin{align}\label{eq:EFEST}
    G_{ab}[\g_{cd}]=T_{ab}^{\scal}[\varphi]+\alpha T_{ab}^\c[\g_{cd},\varphi] +T_{ab}^{\m}[\g_{cd},\varphi,\Psi]\ ,
\end{align}
where $G_{ab}[\g_{cd}]$ is the Einstein operator, 
\begin{align}
    T^{(i)}_{ab}=-\frac{8\pi}{\sqrt{-\g}}\frac{\delta S_{(i)}}{\delta g^{ab}}\ ,
\end{align}
where $i\in \{\c,\m\}$, and
\begin{align}\label{eq:Tscal}
    T_{ab}^{\scal}[\varphi]&=\frac{1}{2}\partial_a\varphi \partial_b \varphi -\frac{1}{4}\g_{ab}\partial_c\varphi\partial^c\varphi\ .
\end{align}

To derive the scalar field equation one varies Eq.~\eqref{eq:action} with respect to $\varphi$, yielding
\begin{align}\label{eq:scalarwave-full}
    \Box_{\g} \varphi = \alpha\Sigma_\c+\Sigma\,,
\end{align}
where $\Box_{\g}  = \g^{ab} \nabla^\g_a\nabla^\g_b$, $\nabla^\g_a$ is the covariant derivative associated with the metric $\g_{ab}$, and
\begin{align}\label{eq:scalarchargedensitym}
    \Sigma=-\frac{16\pi}{\sqrt{-\g}}\frac{\delta S_\m}{\delta\varphi},\qquad  \Sigma_\c=-\frac{16\pi}{\sqrt{-\g}}\frac{\delta S_\c}{\delta\varphi}.
\end{align}

\subsection{The primary}\label{sec:blackhole}

We assume that the primary is a black hole. 
The action $S_0$ is covered by no-hair theorems \cite{Hawking:1972qk,Sotiriou:2011dz}, so any scalar hair would have to be introduced by terms in $S_\c$. For shift-symmetric theories, however, $S_\c$ is also covered by the no-hair theorem, for static and spherically symmetric \cite{Hui:2012qt} and for slowly rotating \cite{Sotiriou:2014pfa} asymptotically flat black holes. The only interaction that evades this theorem is a linear coupling between $\varphi$ and the Gauss-Bonnet invariant ${\cal G}\equiv R_{abcd} R^{abcd}-4 R_{ab} R^{ab}+R^2$ \cite{Sotiriou:2013qea}. Adding the term $\alpha_{\rm GB} \,\varphi \,{\cal G}$ to $S_0$ introduces a non-constant scalar field to black holes, described by the black hole having a scalar charge; however, the scalar charge (which can be thought of as the scalar monopole) is not an independent parameter, instead, the scalar charge is fixed by a regularity condition on the horizon and is determined by the mass and spin of the black hole and $\alpha_{\rm GB}$ \cite{Sotiriou:2013qea,Sotiriou:2014pfa,Thaalba:2022bnt} (see also \cite{Kanti:1995vq} for earlier work without shift symmetry). The charge per unit mass $d$ scales as $\alpha_{\rm GB}/M^2$ in geometric units~\cite{Sotiriou:2014pfa}. 

Adding additional shift-symmetric interactions will change the scalar configuration \cite{Thaalba:2023fmq}, but the regularity conditions that fix the charge persists. The charge per unit mass is then given by an integral over the horizon ${\cal H}$,
\cite{Saravani:2019xwx}
\begin{align}
    \label{chargescaling}
    d=\frac{\alpha_{\rm GB}}{4\pi M}\int_{\cal H} n^a {\cal G}_a \ d\Omega
\end{align}
where $n^a$ is the horizon generator and ${\cal G}=\nabla_a {\cal G}^a$. 
This implies that any terms in $S_\c$ other than the linear coupling with ${\cal G}$, controlled by a coupling $\alpha_i$,  contribute to $d$ with a factor of $\alpha_{\rm GB}\alpha_i$. 

\subsection{The secondary}\label{sec:secondary}

To initially define the matter action for  \eqref{eq:action}, which describes the secondary body in the EMRI, we use the conventional \textit{skeletonized} approach\footnote{The skeletonized formalism was first developed for electromagnetism and gravity~\cite{mathisson2010republication} and has previously been extended to scalar-tensor theories with multiple fields~\cite{damour1992tensor}.} \cite{damour1992tensor}.
The skeletonized description of a compact object replaces the matter action, $S_\m$ (assuming no other matter fields are present), with a point particle action $S_\p$. Ref.~\cite{damour1992tensor} presented a point-particle action for a massive, scalar-charged, compact object:
\begin{align}\label{eq:skeleton1}
    S_\p=-\int_\gamma m[\varphi] ds = - \int_\gamma m[\varphi] \sqrt{\g_{ab} \mathbf{u}^a\mathbf{u}^b}d\tau,
\end{align}
where $\gamma$ is the worldline of the compact object, 
$\mathbf{u}^\alpha=\frac{dz^\alpha}{d\tau}$ is the four-velocity of the compact object in $\g_{ab}$ and $\tau$ is the proper time in $\g_{ab}$. Eq.~\eqref{eq:skeleton1} introduces 
a mass function $m[\varphi]$, which depends on the scalar field, generating the scalar charge. We will show that Eq.~\eqref{eq:skeleton1} is sufficient for deriving the linear field equations but encounters issues beyond linear order.

With our point-particle action in hand, we can now derive the stress-energy tensor and scalar charge density that will appear in the field equations.  Varying Eq.~\eqref{eq:skeleton1} with respect to the metric, $\g_{ab}$, yields 
\begin{align}\label{eq:Tabp}
    T_\m^{ab} = 8\pi \int_\gamma m[\varphi]\frac{\delta^{4}[x^\mu-z_\p^\mu[\tau]]}{\sqrt{-\mathbf{g}}} \mathbf{u}^a \mathbf{u}^b d\tau,
\end{align}
the scalar charged point particle stress-energy tensor.
Varying Eq.~\eqref{eq:skeleton1} with respect to the  scalar field, $\varphi$, yields 
\begin{align}
    \Sigma=16\pi \int_\gamma m'[\varphi]\frac{\delta^{4}[x^\mu-z_\p^\mu[\tau]]}{\sqrt{-\mathbf{g}}}d\tau,\label{eq:T}
\end{align}
the point particle scalar-charge density. 


\section{Perturbative expansion}\label{sec:SSF}

The contribution $\alpha$ makes to $\g_{ab}$ and $\varphi$ must be dimensionless as $\g_{ab}$ and $\varphi$ are dimensionless. As $M$ is the only length scale associated with the background spacetime, $\alpha$ must be accompanied by a factor of $\frac{1}{M^n}$ for the leading-order contribution. We introduce the dimensionless coupling 
\begin{align}\label{eq:zeta}
    \zeta=\frac{\alpha}{M^n} .
\end{align}
This can be expressed in terms of the mass ratio $\varepsilon$ as
\begin{align}\label{eq:zetaepsilon}
    \zeta=\frac{\alpha}{M^n}=\varepsilon^n\frac{\alpha}{\mu^n} \ .
\end{align}
$\zeta$ represents the non-minimal coupling perturbation parameter for $\g_{ab}$ and $\varphi$. 

If we assume that solutions of the field equations are continuously connected to GR solutions as $\alpha \to 0$, our earlier assumptions for $S_\c$, that $[\alpha]=[{\rm length}]^n$, with $n\geq 2$, and the expression for $d$ in Eq.~\eqref{chargescaling}, imply that deviations from GR are controlled by $\zeta$. In particular, so long as the length-(energy)-scale associated with a particular coupling  $\alpha_i$ of a term in $S_\c$ ($\alpha$ denotes them collectively) is not significantly larger (smaller) than the scale associated with $\alpha_{\rm GB}$, $d\propto \alpha_{\rm GB}/M^2$ to leading order in $M^{-1}$. Existing  constraints inferred from astrophysical 
observations imply that 
$\frac{\alpha}{\mu^n}=\O(1)$ or smaller \cite{nair2019fundamental}, as $\mu$ corresponds to solar mass bodies. Therefore, the mass-ratio, $\varepsilon$, the natural 
bookkeeping parameter for the self-force approach, can be used as the sole perturbative parameter for the problem at hand. That is, conservatively,  $\zeta \sim \varepsilon^n$.


To build our formalism, we will consistently expand the 
field's equations \eqref{eq:EFEST}-\eqref{eq:scalarwave-full}, as well as the metric and the scalar field, 
up to the second order in $\varepsilon$. An expansion of a generic tensor, ${\bf A}$, takes the form,
\begin{equation}\label{eq:genericexp}
{\bf A}={\bf A}^{(0)}+\varepsilon {\bf A}^{(1)}+
\varepsilon^2 {\bf A}^{(2)}+\mathcal{O}(\varepsilon^3)\ .
\end{equation}

The metric expansion is written as
\begin{align}\label{eq:gExpansion}  \g_{ab}=g_{ab}+\varepsilon h^{(1)}_{ab}+\varepsilon^2h^{(2)}_{ab}+\O( \varepsilon^3),
\end{align}
where $g_{ab}$ is the background metric (which we take to be the Kerr metric) and $h^{(n)}_{ab}$ are the metric perturbations. The background metric is used to raise and lower the indices of all tensors and tensor perturbations, such as $h^{(n)}_{ab}$. We label all the tensor perturbations with a subscript or superscript number in brackets, which denotes the order in $\varepsilon$ of the perturbation. In practice, the $\varepsilon$ dependence is implicit in the labelled tensors, and the explicit factors of $\varepsilon$ in Eq.~\eqref{eq:gExpansion} are used as a counting parameter. That is, $\varepsilon$ is set to $1$ before computing calculations. 

The scalar field expansion takes the form
\begin{align}\label{eq:phiexpansion}
\varphi=\varphi^{(0)}+\varepsilon\varphi^{(1)}+\varepsilon^2\varphi^{(2)}+\O(\varepsilon^3)\ .
\end{align}
Note that $\varphi^{(0)}$ corresponds to the contribution from the action $S_0$ for an isolated black hole, which is covered by no-hair theorems. Hence, $\varphi^{(0)}$ is constant and can be set to zero by 
a constant shift without loss of generality. 

Our procedure requires a specific treatment for 
the mass function $m[\varphi]$ that appears 
in the secondary's stress-energy tensor \eqref{eq:effectiveTp}. 
To this aim we expand $m[\varphi]$ as:
\begin{align}\label{eq:mexpansion}
     m[\varphi]= m_{[0]}+m_{[1]} \varphi + m_{[2]} \varphi^2 + \O[\varphi^3],
\end{align}
where $m_{[0]}$, $m_{[1]}$ and $m_{[2]}$ are 
constant coefficients. 
In our setup, $m_{[0]}=\mu$; $\mu$ can be considered as the secondary mass in GR (that is, for $\varphi=0$). Note $\mu$ is not the total mass of the secondary in scalar-tensor theories of gravity as $m_{[1]}$ and $m_{[2]}$ can contribute to the mass when $\varphi\neq0$, as we will show. We assume
$m_{[1]}$ and $m_{[2]}$ have the same (stellar 
mass) scale of $m_{[0]}$; that is $m_{[0,1,2]}/M=\mathcal{O}(\varepsilon)$.

We can derive the explicit expression of $m[\varphi]$ 
up to the second order in $\varepsilon$ by replacing 
\eqref{eq:phiexpansion} 
(with $\varphi^{(0)}=0$) into Eq.~\eqref{eq:mexpansion}, obtaining:
\begin{align}\label{eq:mexpansionq}
    m[\varphi]&=\mu +\varepsilon m_{[1]}\varphi_{(1)}+\varepsilon^2 
    (m_{[2]} \varphi_{(1)}^2+m_{[1]}\varphi_{(2)})\ .
\end{align}
\subsection{Field regularization}
We now return to the matter action and show how the conventional \textit{skeletonized} point particle action is problematic in the self-force context. Equation~\eqref{eq:skeleton1} poses a typical problem within self-force: near the worldline ($\gamma$), the singular nature of the metric and scalar field makes 
$\g_{ab}$ and $m[\varphi]$ ill-defined, whereas the four-velocity is only defined on the worldline 
\cite{dirac1938classical}. Hence, Eq.~\eqref{eq:skeleton1} is ill-defined.

To solve this problem and extend Eq.~\eqref{eq:skeleton1} 
beyond linear order, we assume the existence of a \textit{singular} 
(${\cal S}$) and \textit{regular} (${\cal R}$) split of 
the metric and scalar field perturbations~\cite{detwhit2003, zimmerman2015gravitational}: $h^{(n)}_{ab}=h^{(n)\S}_{ab} +h^{(n)\R}_{ab}$, $\varphi^{(n)}=\varphi^{(n)\S} + \varphi^{(n)\R}$. A motivation for this assumption is such a split exists in the decoupled case~\cite{detwhit2003}. We also define 
\begin{align}\label{eq:regular}
h^{\R}_{ab}=\sum_{n=1} \varepsilon^n h^{\R(n)}_{ab}\quad,\quad
\varphi^{\R}=\sum_{n=1} \varepsilon^n \varphi^{(n)}_\R.
\end{align}
$h^{\R}_{ab}$ and $\varphi^{\R}$ identify
the regular part of the field perturbations (the part that generates the self-force) \cite{dirac1938classical}. Ref.~\cite{detwhit2003} provides an extensive analysis of the singular-regular decomposition to linear order, which has been generalised to the non-linear regime in GR~\cite{harte2012mechanics, pound2012field, harte2015motion}.

The decomposition into singular and regular pieces 
of the metric perturbations has been studied extensively in GR 
in \cite{detwhit2003}. Conventionally, $G_{ab}^{(1)}[h^{(1)\S}_{cd}]=8\pi T^{\rm m (1)}_{ab}$ and $G_{ab}^{(1)}[h^{(1)\R}_{cd}]=0$ are satisfied. 
This definition does not fully fix the field; additionally, $h^{\S}_{ab}$ is chosen to depend only on the 
instantaneous state and position of the particle, and 
$h^{\R}_{ab}$ on the compact objects' 
causal past~\cite{detwhit2003}. We expect similar definitions to hold for the metric perturbation and scalar field in our formalism.

With Eqns.~\eqref{eq:regular} in hand we define an 
effective metric and scalar field:
\begin{align}
\tilde{g}_{ab}=g_{ab}+h^{\R}_{ab}\quad, \quad
    \tilde{\varphi}=\varphi^{\R}\ .\label{eq:tildephiexpansion}
\end{align} 
First and second-order self-force calculations in GR 
have found that compact objects move as a test body of 
the effective 
metric \cite{misata1997, quwa1997, pound2012second, gralla2012}. 
%
%

Replacing $\g_{ab}$ and $\phi$ with $\tilde{g}_{ab}$ and $\tilde{\varphi}$ 
in Eq.~\eqref{eq:skeleton1}, we obtain our \textit{effective} point-particle action of a scalar-charged compact object:
\begin{align}\label{eq:effectiveskeleton}
    S_\p=-\int_{\gamma} m[\tilde{\varphi}] d\tilde{s} = - \int_{\gamma} m[\tilde{\varphi}] \sqrt{\tilde{g}_{ab} \tilde{u}^a\tilde{u}^b}d\tilde{\tau},
\end{align}
where $\tilde\tau$ is the proper time in the effective 
spacetime and $\tilde{u}^\alpha=dz^\alpha/d\tilde{\tau}$.  
Equation~\eqref{eq:effectiveskeleton} represents our ansatz for the point particle action. 
For $\tilde{\varphi}=0$ it reduces to the point particle 
action in GR, Eq.~\eqref{eq:effectiveskeletonGR}, which we show to be consistent with 
self-force calculations up to second-order~\cite{Upton, pound2012second, gralla2012} in App.~\ref{app:SFGR}.

By varying the action~\eqref{eq:effectiveskeleton} with respect to the effective metric, $\tilde{g}_{ab}$, yields 
\begin{align}\label{eq:effectiveTp}
    T_\m^{ab} = 8\pi \int_\gamma m[\tilde\varphi]\frac{\delta^{4}[x^\mu-z_\p^\mu[\tilde\tau]]}{\sqrt{-\tilde g}} \tilde{u}^a \tilde{u}^b d\tilde\tau,
\end{align}
varying with respect to the effective scalar field, $\tilde \varphi$, yields 
\begin{align}
    \Sigma=16\pi \int_\gamma m'[\tilde\varphi]\frac{\delta^{4}[x^\mu-z_\p^\mu[\tilde\tau]]}{\sqrt{-\tilde{g}}}d\tilde\tau,\label{eq:Ttilde}
\end{align}
which replace Eqs.~\eqref{eq:effectiveTp} and \eqref{eq:Ttilde} respectively. Eqs.~\eqref{eq:effectiveTp} and~\eqref{eq:Ttilde} inform us that $m[\tilde\varphi]$ and $m^\prime[\tilde\varphi]$ are the Bondi mass and scalar charge of the secondary compact object respectively.

We remark that our approach is motivated by 
Ref.~\cite{detweiler2012second-order}, which applied 
the effective metric approach to the point-particle stress-energy of a massive compact object in GR. 
Their main result was a conjecture for the second-order stress-energy tensor of a compact object in self-force in GR. This conjecture was later proven to hold in regular gauges by Ref.~\cite{Upton}. 

\subsection{The equation of motion for the secondary}

Using our point particle action, Eq.~\eqref{eq:effectiveskeleton}, 
we can also derive the equation of motion of the compact object. 
We follow the approach of~\cite{zimmerman2015gravitational}, 
varying the whole action~\eqref{eq:action} with respect to the 
body's path, $x^\mu\rightarrow x^\mu +\delta x^\mu$. This 
yields:
\begin{align}
    \delta_{x^\mu} S_{\p} &=\int_\gamma \bigg[-(\tilde g^\alpha_{\ \mu}+\tilde u^\alpha \tilde u_\mu)\frac{\partial m[\varphi]}{\partial \varphi} \frac{\partial \varphi}{\partial x^\alpha}\delta x^\mu \notag \\
    &+ \delta x^\mu m[\varphi] \Big( \tilde \Gamma_{\mu\alpha\nu} \tilde u^\alpha \tilde u^\nu - \tilde g_{\mu\nu}\frac{d^2x^\nu}{d\tilde \tau^2} \Big)\bigg]d\tilde \tau\  +\order{\varepsilon^3}\ .
\end{align}
Note that the contribution coming from the non-minimal action 
is at least order $\order{\varepsilon^3}$ - based on our earlier assumptions that $\zeta=\order{\varepsilon^2}$ and $S_\c$ contains at least one copy of $\varphi$, which is itself order $\varepsilon$. 

Requiring stationarity under first-order 
variations yields
\begin{align}\label{eq:effectivea}
   m[\tilde \varphi] \tilde{a}^a=m^\prime[\tilde \varphi](\tilde g^{ab}+\tilde u^a\tilde u^b) \partial_b \tilde\varphi +\order{\varepsilon^3}\ ,
\end{align}
where $\tilde{a}^a=\tilde{u}^b\tilde{\nabla}_b\tilde{u}^a$ 
and $\tilde{\nabla}_b$ is the covariant derivative of the 
effective metric. 
Eq.~\eqref{eq:effectivea} is equivalent to the standard self-force equation of motion 
for a point scalar charge~\cite{quinn2000axiomatic, gralla2010motion, poisson2011living} but extended to at least second-order. The charge moves as a point-particle being pushed away from geodesic motion in the effective spacetime by a self-force generated by the effective scalar field.

We can also derive evolution equations for the mass and scalar charge of the secondary compact object~\cite{zimmerman2015gravitational}:
\begin{align}
    \frac{dm}{d\tau}= \frac{\partial m}{\partial \tilde \varphi}\frac{\partial \tilde \varphi}{\partial \tau} =m^\prime[\tilde \varphi] u^a\nabla_a \tilde \varphi, \label{eq:mevol}\\
    \frac{dm^\prime}{d\tau}= \frac{\partial m^\prime}{\partial \tilde \varphi}\frac{\partial \tilde \varphi}{\partial \tau} =m^{\prime\prime}[\tilde \varphi] u^a\nabla_a \tilde \varphi.\label{eq:mprimeevol}
\end{align}

Note, Eqns.~\eqref{eq:effectiveTp}-\eqref{eq:Ttilde}, and~\eqref{eq:effectivea} reduce to the GR limit, Eq.~\eqref{eq:effectiveaGR}, when $\tilde \varphi\rightarrow 0$. 
The validity of the equations of motion for black holes and 
self-gravitating extended compact objects in an effective spacetime 
has been assessed in GR up to the second order in 
the self-force expansion \cite{harte2012mechanics, misata1997, quwa1997, pound2012second, gralla2012}. Our results here are analogous 
but in scalar-tensor theories of gravity, up to our assumptions.

\section{Perturbed equations}\label{sec:perteqs}

To isolate first- and second-order contributions of the field's equations~\eqref{eq:EFEST} in orders of $\varepsilon$ we need to expand 
the non-linear behaviour of the Einstein tensor, $G_{ab}[\g_{ab}]$. 
For a generic tensor $g^{(0)}_{ab}+x_{ab}$, the expansion can be expressed as 
\begin{align}\label{eq:Gexpansion}
G_{ab}[x_{ab}]= \delta G_{ab}[x_{ab}]&+\delta^2 G_{ab}[x_{ab},x_{ab}]\notag \\ &+\delta^3 G_{ab}[x_{ab},x_{ab},x_{ab}]+\ldots
\end{align}
where 
\begin{equation}
\delta^n G = \frac{1}{n!}\frac{d^n}{d\lambda^n}G\left[g^{(0)}_{ab}+\lambda x_{ab}\right]\big|_{\lambda=0}\ .
\end{equation}
Replacing the expansion \eqref{eq:genericexp} for the 
metric into Eq.~\eqref{eq:Gexpansion} gives:
\begin{align}\label{eq:expandedEFE}
G_{ab}[\g_{ab}]&=\varepsilon \delta G_{ab}[h^{(1)}_{ab}] \notag \\ & \ \ \ + \varepsilon^2 \Big(\delta G_{ab}[h^{(2)}_{ab}]+\delta^2G_{ab}[h^{(1)}_{ab},h^{(1)}_{ab}] \Big) +\mathcal{O}(\varepsilon^3), \notag \\
&= \varepsilon G^{(1)}_{ab}[\g_{ab}] +  \varepsilon^2 G^{(2)}_{ab}[\g_{ab}] + \mathcal{O}(\varepsilon^3)
\end{align}
where we have used $G_{ab}[g^{(0)}_{ab}]=0$. 

We shall now derive the explicit form for the first- 
and second-order field equations which determine 
the evolution of the metric and scalar perturbations. We also derive the first- and second-order equations of motion of the compact body, which depend on said perturbations. Deriving these equations requires expanding all the quantities defined in the metric ($\g_{ab}$) and effective metric ($\tilde g_{ab}$) as perturbative expansions around the background metric $g^{(0)}_{ab}$. The required expansions are Eqs.~\eqref{eq:phiexpansion},~\eqref{eq:gExpansion},~\eqref{eq:tildephiexpansion}, and~\eqref{eq:effectivea}, plus some non-trivial expansions given in App.~\ref{app:expansiontools}.

\subsection{First order perturbations}\label{sec:firstorder}

Expanding Eq.\eqref{eq:EFEST} to linear order in the mass-ratio, the field equation 
for $h_{ab}^{(1)}$ are given by 
\begin{equation}\label{eq:dGh1}
    G^{(1)}_{ab}=\delta G_{ab}[h^{(1)}_{ab}]= 8\pi\mu \int_\gamma \frac{\delta^{4}[x^\mu-z_\p^\mu[\tau]]}{\sqrt{-g}}u_au_b d\tau\ ,
\end{equation}
which are the same as in GR\footnote{In order for Eq~\eqref{eq:dGh1} to satisfy the Bianchi identity $\delta G_{ab}$ must be gauge fixed (e.g., to the Lorenz gauge) in the self-consistent approximation~\cite{pound2010self}. This would similarly impose restrictions on our action~\eqref{eq:action}. Alternatively, one can adopt the Gralla--Wald approximation~\cite{grallawald2008}, expanding the worldline in powers of $\varepsilon$, but this approach does not admit a multi-scale expansion.}. Hence, the standard methods of metric 
reconstruction~\cite{chrzanowski1975vector,cohen1975space,green2020teukolsky, toomani2021new,aksteiner2019new,dolan2022gravitational} hold.
Note that Eq.~\eqref{eq:dGh1} can be derived using either \eqref{eq:Tabp} or \eqref{eq:effectiveTp} for $T^\m_{ab}$ because this is linear order; that is, the result is equivalent to the metric field equation in Refs.~\cite{maselli2020detecting, Maselli:2021men, barsanti2022extreme}.

The wave equation for the scalar field perturbation 
$\varphi^{(1)}$ can be obtained by expanding 
Eq.~\eqref{eq:scalarwave-full} to linear order, 
\begin{align}\label{eq:first-orderscalarwave}
    \Box \varphi^{(1)} &= 16\pi \int_\gamma m_{[1]}\frac{\delta^{(4)}[x^\mu-z_\p^\mu[\tau]]}{\sqrt{-g}}d\tau\ , 
\end{align}
where $\Box = g^{ab}_{(0)} \nabla_a\nabla_b$ and $\nabla_a$ 
is the covariant derivative associated with the background 
metric $g^{(0)}_{ab}$. Similarly, using either \eqref{eq:T} or \eqref{eq:Ttilde} for $\Sigma$ is viable at this order.

Eqs.~\eqref{eq:dGh1} and~\eqref{eq:first-orderscalarwave} are equivalent to the decoupled first-order field equations. Hence, we can use the regular and singular field split for $h^{(1)}_{ab}$ and $\varphi^{(1)}$ developed in Ref.~\cite{detwhit2003}.

As shown in \cite{maselli2020detecting, Maselli:2021men, Barsanti:2022ana}, at first-order in $\varepsilon$ the 
mass parameter $m_{[1]}$ can be interpreted in 
terms of the scalar charge of the secondary. 
Expanding the scalar field in a buffer region inside 
the world-tube containing the compact object, such 
that $\mu\ll\hat{r}\ll M$, with $\hat{r}$ being the distance from the worldline, gives
%
\begin{align}\label{eq:phi-r-expansion}
    \varphi^{(1)}=\frac{\mu d}{\hat r} + \O\Big(\frac{\mu^2}{\hat{r}^2}\Big) ,
\end{align}
where $d$ is the dimensionless scalar charge.
Replacing Eq.~\eqref{eq:first-orderscalarwave} 
into the field's equation 
\eqref{eq:phi-r-expansion}, and matching the 
solution in the buffer zone 
leads to 
\begin{align}\label{eq:m1}
    m_{[1]}=-\frac{\mu d}{4}\ .
\end{align}

Equations~\eqref{eq:dGh1}-\eqref{eq:first-orderscalarwave} have been numerically integrated in \cite{maselli2020detecting,Barsanti:2022ana} 
for binaries on circular and eccentric orbits, to compute 
the leading order dissipative correction to the energy and 
the angular momentum gravitational wave fluxes. These results are equivalent the dissipative piece of the first order self force and are sufficient 
for evolving the binary dynamics to adiabatic accuracy.

The solution for $h^{(1)}_{ab}$ and 
$\varphi^{(1)}$ can be used to compute the full (dissipative plus conservative) first-order self-force using the first-order equation of motion. Expanding Eq.~\eqref{eq:effectivea} gives the first-order equations of 
motion\footnote{Note, Eq.~\eqref{eq:a1} cannot be derived directly from Eq.~\eqref{eq:skeleton1} because of the lack of regularisation; that is, the effective action, Eq.~\eqref{eq:effectiveskeleton}, is required. The expansion also requires Eq.~\eqref{eq:dtautildeexpanaion}}
\begin{align}\label{eq:a1}
    a^a_{(1)}= a^{a}_{(1)\grav}+a^{a}_{(1)\scal}\ ,
\end{align}
where the gravitational and the scalar components 
are given by:
\begin{equation}\label{eq:a1grav}
     a^a_{(1)\grav} =  -\frac{1}{2} (g^{ab} +u^au^b)(2h^{(1)\R}_{bd;e}-h^{(1)\R}_{de;b})u^d u^e\ ,
\end{equation}
which is identical to Eq.~\eqref{eq:a1gravGR}, and 
\begin{equation}\label{eq:a1scal}
    a_{(1)\scal}^a =m_{[1]} (g^{ab} + u^a u^b)\nabla_b \varphi^{(1)}_\R \ .
\end{equation}



Calculations for both $a^{(1)}_{\grav}$ and $a^{a}_{(1)\scal}$ 
have been carried out for 
generic orbits in the Kerr background in the literature~\cite{van2015metric,van2016gravitational, van2018gravitational,heffernan2022regularization}. These results can be exploited to derive 
the full first-order self-force in scalar-tensor theories, 
which includes conservative corrections to an 
EMRI's evolution.

\subsection{Second order perturbations}\label{sec:secorder}





Using our expansion of Eq.\eqref{eq:EFEST} (including the expansions in App.~\ref{app:expansiontools}), we can express the 
field's equations for the second-order metric perturbation 
\begin{align}
    & \delta G_{ab}[h^{(2)}_{cd}]=-\delta^2G_{ab}[h^{(1)}_{cd},h^{(1)}_{cd}]+\frac{1}{2}\partial_a\varphi^{(1)} \partial_b \varphi^{(1)}\notag \\
    &-\frac{g_{ab}}{4}\partial_c\varphi^{(1)}\partial^c\varphi^{(1)} + 4\pi \int_\gamma \frac{\delta^{4}[x^\mu-z_\p^\mu[\tau]]}{\sqrt{-g}}\bigg[2m_{[1]}\varphi^{(1)}_\R u_au_b \notag \\
    &+ \mu\big( 4h^{\R(1)}_{ac}u^cu_b-u_a u_b (g^{cd}_{(0)} - u^{c}u^{d})h^{\R(1)}_{cd}\big) \bigg] d\tau \ .\label{eq:dGh2}
\end{align}
Note, $G^{(2)}_{ab}=\delta G_{ab}[h^{(2)}_{cd}]+\delta^2G_{ab}[h^{(1)}_{cd},h^{(1)}_{cd}]$.
For $m_{[1]}=0$ and $\varphi^{(1)}_\R=0$ 
the right hand side of Eq.~\eqref{eq:dGh2} reduces 
to the GR form, Eq.~\eqref{eq:dGh2GR}. 

Expanding the scalar field equation, Eq.~\eqref{eq:scalarwave-full} (including the expansions in App.~\ref{app:expansiontools}),  to second-order, we obtain:
\begin{align}
    \Box \varphi^{(2)} &=  -\frac{8\pi\alpha^{(2)}}{\sqrt{-g}}\G^{(0)} - h^{ab}_{(1)} \nabla_a\nabla_b\varphi^{(1)} -(\nabla^a h^{(1)}_{ab})\nabla^b\varphi^{(1)} \notag \\
    & \ \ \ +\frac{1}{2} (\nabla^b h_{(1)}) \nabla_b \varphi^{(1)}+16\pi  \int_\gamma \Big[ m_{[2]}\varphi^{(1)}_R  \notag \\
    & \ \ \  -\frac{1}{2}  m_{[1]} (g^{ab}+u^au^b)h_{ab}^{R(1)} \Big] \frac{\delta^{(4)}[x^\mu-z_\p^\mu[\tau]]}{\sqrt{-g}} d\tau  . \label{eq:second-orderscalarwave}
\end{align}
where $h_{(1)}=g^{ab}_{(0)}h^{(1)}_{ab}$\footnote{Eq.~\eqref{eq:dGh2} and~\eqref{eq:second-orderscalarwave} are both problematic in the sense they are not well defined as distributions. This is due to both equations containing products of quantities that are singular $h^{(1)}_{ab}$ and $\varphi^{(1)}$ on the worldline. However, it is possible to construct well-defined equations by using a puncture scheme or regularisation prescription (such as Hadamard finite part~\cite{hadamard1932probleme}) following Ref.~\cite{Upton}. In calculating the second-order puncture~\cite{pound2014practical}, one will also define the existence of an appropriate singular and regular splits of $h^{(2)}_{ab}$ and $\varphi^{(2)}$, which we leave for future work.}. As discussed earlier, $\zeta S_\c$ is at least  $\order{\varepsilon^3}$, as $\zeta$ is at least order $\varepsilon^2$ and $S_\c$ contains at least one copy of $\varphi$. Due to this, there are no contributions from $S_\c$ to Eq.~\eqref{eq:dGh2} and the only contribution to Eq.~\eqref{eq:second-orderscalarwave} comes from a linear coupling between $\varphi$ and ${\cal G}$ \cite{Sotiriou:2013qea}. That is
\begin{align}
T_{ab}^\c&=\O(\varepsilon^3), \\
    \Sigma_\c&= \varepsilon^2\frac{8\pi\alpha^{(2)}}{\sqrt{-g}}\G^{(0)} +\O(\varepsilon^3).\label{eq:dSc}
\end{align}
For the leading order contribution in Eq.~\eqref{eq:dSc}, $\alpha$ has dimension $[{\rm length}]^2$, so we label $\alpha\rightarrow\alpha^{(2)}$ as $\alpha \G^{(0)}=\order{\varepsilon^2}$. In a Kerr background
\begin{align}
    \G^{(0)}=24M^2\big((r-ia\cos[\theta])^{-6}+(r+ia\cos[\theta])^{-6}\big).
\end{align}

The expansion of the equation of motion, Eq.~\eqref{eq:effectivea} (using Eq.~\eqref{eq:dtautildeexpanaion}), to second-order gives
\begin{align}
    a^a_{(2)}= a^{a}_{(2)\grav}+a^{a}_{(2)\scal}\ ,\label{eq:a2}
\end{align}
with 
\begin{align}\label{eq:a2grav}
    & a^a_{(2)\grav} = -\frac{1}{2} \Big[(g^{ab}+u^au^b)(2h^{(2)\mathcal{R}}_{bd;e}-h^{(2)\mathcal{R}}_{de;b})  \notag \\ & \ \ \ \ \ - (g^{ab}+u^au^b)h_{b(1)\mathcal{R}}^{\  c}(2h^{(1)\mathcal{R}}_{cd;e}-h^{(1)\mathcal{R}}_{de;c}) \Big] u^d u^e \ ,
\end{align}
which is identical to Eq.~\eqref{eq:a2gravGR}, and 
\begin{align}
    \mu a_{(2)\scal}^a &=  (g_{(0)}^{ab} + u^a u^b)\Big(m_{[1]}\nabla_b  \varphi^{(2)}_{\R} \notag \\ 
    & \ \ \ +2m_{[2]}\varphi^{(1)}_{\R}\nabla_b  \varphi^{(1)}_{\R} - \frac{m_{[1]}^2}{\mu}\varphi^{(1)}_{\mathcal{R}}\nabla_b  \varphi^{(1)}_{\mathcal{R}} \Big) \notag \\ 
    & \ \ \ + m_{[1]}\big(h_{cd}^{\mathcal{R}} u^cu^du^au^b -h^{ab}_{\mathcal{R}} \big)\nabla_b \varphi^{(1)}_{\mathcal{R}}.\label{eq:a2scal}
\end{align}
Eq.~\eqref{eq:a2grav} is equivalent to the second-order 
self-force in GR, Eq.~\eqref{eq:a2gravGR}\footnote{The second-order self-force equations, Eqs.~\eqref{eq:a2grav} and \eqref{eq:a2scal} (and similarly calculating $h^{(2)\mathcal{R}}_{ab}$ and $\varphi^{(2)}_{\mathcal{R}}$) may be 
unnecessary if flux balance laws can be derived to extract the dissipative piece of the second-order self-force directly from $h^{(2)}_{ab}$ and $\varphi^{(2)}$.}.


The interpretation of $m_{[2]}$ is similar to $m_{[1]}$ except its contribution to the scalar charge is suppressed by an order in $\varepsilon$ as there is further coupling to the scalar field $\varphi^{(1)}_{\R}$. Examining the $m_{[2]}$ piece in Eq.~\eqref{eq:second-orderscalarwave} we see it will also generates a term in $\varphi^{(2)}$ equivalent to Eq.~\eqref{eq:phi-r-expansion}. Note, the other delta function term (which is coupled to $h^{\R(1)}_{ab}$) in Eq.~\eqref{eq:second-orderscalarwave} will also contribute a term equivalent to Eq.~\eqref{eq:phi-r-expansion}. Therefore,
\begin{align}
    \varepsilon\mu d&=-4\bigg(\varepsilon m_{[1]} \notag \\
    & \ \ \ +\varepsilon^2\Big(m_{[2]}\varphi^{(1)}_R -\frac{1}{2}  m_{[1]} (g^{ab}+u^au^b)h_{ab}^{\R(1)} \Big)\bigg).
\end{align}
That is, $m_{[2]}$, which we call 
hereafter \textit{"charge coupling"}, 
and the coupling to $h_{ab}^{R(1)}$, provide an $\mathcal{O}(\varepsilon)$ correction to the scalar charge $d$. We can expand $d$,
\begin{align}
    d=d^{(0)}+\varepsilon d^{(1)} +\order{\varepsilon^2}.
\end{align}
We re-define Eq.\eqref{eq:m1} as
\begin{align}\label{eq:m1b}
    m_{[1]}=-\frac{\mu d^{(0)}}{4} ,
\end{align}
and define
\begin{align}\label{eq:d1}
         d^{(1)}&=-\frac{4}{\mu}\Big(m_{[2]}\varphi^{(1)}_R -\frac{1}{2}  m_{[1]} (g^{ab}+u^au^b)h_{ab}^{\R(1)} \Big).
\end{align}

Interestingly, the only piece in the second-order equations that does not derive from a minimally coupled scalar field to the metric is the Gauss-Bonnet invariant term in Eq.~\eqref{eq:second-orderscalarwave}, which is stationary in Kerr spacetime. Its effect on $\varphi^{(2)}$ in Eq.~\eqref{eq:second-orderscalarwave} is then also stationary and, hence, it will not affect the dissipative piece of the second-order self-force. It can, therefore, be neglected for first-post adiabatic accurate modelling. Hence, our formalism is independent of the choice of scalar-tensor theories which obey our assumptions.

In general (including in the GR limit), the most challenging 
term to solve in Eqs.~\eqref{eq:dGh2} is 
computing the mode decomposition of $\delta^2G_{ab}[h^{(1)}_{ab},h^{(1)}_{ab}]$ near the worldline. This problem derives from $\delta^2G_{ab}[h^{(1)}_{ab},h^{(1)}_{ab}]$ being quadratic in the first-order metric perturbation, which is 
singular on $\gamma$. Ref.~\cite{miller2016second} describes and addresses this problem in detail by splitting 
$\delta^2G_{ab}[h^{(1)}_{ab},h^{(1)}_{ab}]$ into three 
pieces: (i) the regular term 
$\delta^2G_{ab}[h^{(1)\R}_{ab},h^{(1)\R}_{ab}]$, 
(ii) the mildly singular term, $\delta^2G_{ab}[h^{(1)\S}_{ab},h^{(1)\R}_{ab}]$ 
computed by casting $h^{(1)\S}_{ab}$ and $h^{(1)\R}_{ab}$ 
as a sum of modes, and (iii) the very singular term, $\delta^2G_{ab}[h^{(1)\S}_{ab},h^{(1)\S}_{ab}]$, which can be calculated using a 4-dimensional expression for $h^{(1)\S}_{ab}$ associated with the mode expansion of $h^{(1)\S}_{ab}$. While this approach has been recently used to calculate the first-post adiabatic quasi-circular inspiral of a Schwarzschild binary system~\cite{pound2020second, wardell2021gravitational, warburton2021gravitational}, the method remains computationally expensive and highly technical. Overcoming this issue remains a major obstacle for self-force waveform modelling 
in GR.

One may have expected this problem to be exacerbated in scalar-tensor theories of gravity as introducing additional degrees of freedom may have resulted in $\delta^2G_{ab}[h^{(1)}_{ab},h^{(1)}_{ab}]$ being different, for different theories. However, the decoupling of scales on which our approach builds provides key simplifications. As the first-order metric perturbation, $h^{(1)}_{ab}$, 
is the same in GR and in all the theories of gravity 
specified by the action~\eqref{eq:action}, the 
expression for $\delta^2G_{ab}[h^{(1)}_{ab},h^{(1)}_{ab}]$ 
remain the same.
Therefore, we can use the same $\delta^2G_{ab}[h^{(1)}_{ab},h^{(1)}_{ab}]$ constructed in GR, 
significantly reducing the computational burden 
of our formalism.

Nonetheless, additional singular terms appear 
at the second order, as quadratic functions of the scalar field contribute in Eq.~\eqref{eq:dGh2}. 
Additionally, mixed products between the metric and the 
scalar linear perturbations appear in
Eq.~\eqref{eq:second-orderscalarwave}. 
These terms are similarly challenging to calculate as they contain products of singular functions on the worldline. However, their mode decomposition can be computed 
adopting the same method 
used in \cite{miller2016second}.
Remarkably, since $\varphi^{(1)}$ 
is independent of the specific theory of 
gravity, up to an amplitude rescaling 
given by the scalar charge, all these 
contributions are scalar-tensor theory invariant.

\section{Two-timescale expansion}\label{sec:two-timescale}
The two-timescale expansion is an example of a multi-scale expansion~\cite{kevorkian1996method}. In the EMRI context, Ref.~\cite{hinderer2008} used the two-timescale expansion to argue that first-post adiabatic models were necessary to model inspirals accurately. Since then, there has been growing interest in applying the two-timescale approximation to the EMRI problem~\cite{miller2020two, wardellpoundreview2021, durkan2022slow, mathews2022self}. A two-timescale expansion was implemented to produce the first-post-adiabatic waveform models for Schwarzschild black holes in a quasi-circular inspiral in GR ~\cite{pound2020second, wardell2021gravitational, warburton2021gravitational}. Here, we apply the two-timescale approximation to our formalism similarly. We show that the resemblance to the two-timescale framework in GR allows calculations in scalar-tensor theories of gravity to require only supplementary terms. 

As discussed in Appendix~\ref{sec:AA1}, EMRI dynamics allow us to identify two distinct timescales~\cite{hinderer2008, miller2020two}: (i) the \textit{fast-timescale} over which the orbital phases evolve and (ii) the \textit{slow-timescale} that dictates the change of the orbital frequencies and of the physical parameters of the system. Due to the near periodicity of each EMRI orbit, the time evolution on the fast-timescale is effectively periodic. In contrast, the slow-time evolution is non-trivial but contributes beyond the leading order (as $\tilde t=\O(\frac{1}{\varepsilon })$)~\cite{miller2020two}. 


The position of the compact object at any given time can be represented by the three orbital phases,
\begin{align}
    \phi_i:=\{\phi_r,\phi_\theta,\phi_\phi\}.
\end{align}
The evolution of the phases, $\phi_i$, can be expressed in terms of their respective frequencies 
$\Omega_i:=\{\Omega_r,\Omega_\theta,\Omega_\phi\}$,
\begin{align}\label{eq:phi}
    \phi_i=\int\Omega_i[\tilde t] dt,
\end{align}
where $\Omega_i$ evolves on the slow-time $\tilde t$. 
We assume $\Omega_i$ is approximately constant on the fast-timescale. This assumption is valid because the background spacetime is Kerr and geodesics in Kerr are triperiodic. An appropriate choice of \textit{fast-times} are the phases, $\phi_i$, as they evolve on the orbital timescale. 

The frequencies, $\Omega_i$, can be expressed in terms of the three constants of motion of Kerr geodesics: 
energy, angular momentum and the Carter constant,
$J_i := \{E,J_z, Q\}$~\cite{barackpound18}. That is,
\begin{align}
    \Omega_i:=\Omega_i[J_i].
\end{align}
As the compact object does not remain on a geodesic over the course of the inspiral, the three constants of motion evolve over the slow-timescale. Their evolution can be computed through the self-force. In turn, the evolution of the frequencies of motion, $\Omega_i$ can be constructed from the self-force,
\begin{align}
    \frac{d \Omega_i}{dt} &= \varepsilon F_{\Omega_i}^{\{0\}}[\Omega_i] + \varepsilon^2 F_{\Omega_i}^{\{1\}}[\Omega_i,\delta M_\mu] + {\cal O}(\varepsilon^3),\label{eq:domega}
\end{align}
where $F_{i}^{\{n\}}$ are $n$th-post adiabatic-order self-force coefficients. Note the adiabatic self-force coefficients only depend on the frequencies. In contrast, the first-post adiabatic coefficients depend on the frequencies and the change in the system's physical parameters $\delta M_\mu$, which we explain next.

Additional physical parameters of the binary evolve on the slow-timescale of an EMRI; one must account for their evolution to achieve first-post adiabatic accurate models. In the GR case, the mass and spin of the primary black hole evolve as gravitational waves pass into the primary black hole horizon. The change in mass and spin are labelled as 
\begin{align}\label{eq:deltaM}
    \delta M_A [\tilde t]:= \{\delta M,\delta J\}.
\end{align} 

We now assess whether additional physical parameters of an EMRI need to be evolved in scalar-tensor theories of gravity. In scalar-tensor theories of gravity, the primary black hole also absorbs scalar radiation. This scalar radiation carries energy and angular momentum, which must be accounted for in the evolution of the $\delta M_A [\tilde t]$. One may question if the scalar field can carry scalar charge into the supermassive black hole. However, in general, scalar charge does not tend to be a free parameter in scalar-tensor theories of gravity. Instead, when present, its value is fixed in terms of the mass and spin of the black hole by regularity conditions on the horizon (see {\em e.g.}~\cite{Kanti:1995vq, Sotiriou:2013qea, Sotiriou:2014pfa, Thaalba:2022bnt}). Hence, we expect the scalar charge of the supermassive black hole to evolve consistently with the evolution of the mass and angular momentum of the black hole. The scaling arguments in Secs.~\ref{sec:blackhole} and ~\ref{sec:SSF}, that multiple orders of $\varepsilon$ suppress the scalar charge of the supermassive black hole, still hold. As the evolution of the mass is an $\order{\varepsilon}$ effect, we can neglect the evolution of the scalar charge of the supermassive black hole as it is a higher-order effect (at least an $\order{\varepsilon^3}$ effect for $n\geq 2$). 

A potential caveat of this argument may be that the expectation that the scalar charge is fixed with respect to the mass and spin is based on the properties of stationary black holes. It is known that relaxing the assumption of stationarity can allow for hair formation in principle \cite{jacobson1999primordial}, but it is reasonable to expect that the charge per unit mass of the primary introduced by the absorption of scalar radiation will be negligible in the present scenario. Further investigation of whether regular (near the horizon) perturbative modes could excite further independent scalar charge degrees of freedom within this multi-scale formalism would test this hypothesis. We can also test this hypothesis against time-domain self-force evolutions, for which the formalism in the main body of this paper is also consistent. Fully nonlinear numerical investigations of a supermassive black hole absorbing scalar radiation arising from an asymmetric binary would also be of interest in this respect.

We next turn our attention to whether the physical characteristics of the secondary evolve on the orbital timescale. For EMRIs in GR, the evolution of the mass of the secondary object is a high-order effect~\cite{mathews2022self}. This is due to the length scales of the orbital dynamics and the radiation wavelength being much larger than the scale of the secondary. This argument extends to scalar-tensor theories of gravity, so we expect the secondary object's parameters $\mu$ and $m_{[1]}$ (related to $\mu$ by Eq.~\eqref{eq:m1b}) to remain constant throughout the inspiral for first-post adiabatic modelling. This is not to say that the mass ($m[\tilde\varphi]$) and scalar charge ($\m^\prime[\tilde\varphi]$) of the secondary remain constant, they evolve via Eqs.~\eqref{eq:mevol} and~\eqref{eq:mprimeevol} respectively. That is, their evolution is determined solely by the evolution of the scalar field, which is accounted for in the two-timescale formalism. Hence, no additional parameters evolve on the slow-timescale in our formalism are required; that is, Eq.~\eqref{eq:deltaM} holds.

The evolution of the EMRI parameters, $\delta M_A$, can also be constructed from the self-force,
\begin{align}
    \frac{d \delta M_A}{dt} &= \varepsilon F_A^{\{1\}}[\Omega_j] + {\cal O}(\varepsilon^2).\label{eq:dM_mu}
\end{align}
More precisely, the evolution of $\delta M$ and $\delta J$ can be calculated from the first-order scalar and gravitational fluxes that pass through the supermassive black hole horizon.

To implement the two-timescale approximation we need to re-express the first- and second-order field equations (Eqs.~\eqref{eq:dGh1},~\eqref{eq:first-orderscalarwave},~\eqref{eq:dGh2}, and~\eqref{eq:second-orderscalarwave}) and the equations of motion (Eqs.~\eqref{eq:a1} and~\eqref{eq:a2}). In the two-timescale approximation, the field variables are expressed as~\cite{miller2020two}
\begin{align}\label{eq:twotimemetpert1}
h^{(1)}_{ab}&= \sum_{p,q,m} h^{(1),\omega_{p,q,m}}_{ab}[\Omega_i,x^i]e^{-ik^i \phi_i}, \\
\varphi^{(1)}&= \sum_{p,q,m} \varphi^{(1),\omega_{p,q,m}}[\Omega_i, x^i]e^{-ik^i \phi_i}, \label{eq:twotimephi1}
\end{align}
and
\begin{align}\label{eq:twotimemetpert2}
h^{(2)}_{ab}&= \sum_{p,q,m} h^{(2),\omega_{p,q,m}}_{ab}[\Omega_i, \delta M_A,x^i]e^{-ik^i \phi_i}, \\
\varphi^{(2)}&= \sum_{p,q,m} \varphi^{(2),\omega_{p,q,m}}[\Omega_i, \delta M_A,x^i]e^{-ik^i \phi_i}, \label{eq:twotimephi2}
\end{align}
where $ k^i:= \{p,q,m\}$, $\omega_{p,q,m}=k^i \Omega_i$, and $p$, $q$, and $m$ are integers to sum over. Eqs.~\eqref{eq:twotimemetpert1},~\eqref{eq:twotimemetpert1},~\eqref{eq:twotimemetpert2}~and~\eqref{eq:twotimephi2} are a discrete Fourier series in terms of the phases of motion, whose coefficients evolve on the slow-timescale. 

The advantage of the two-timescale approximation becomes apparent when you act with a time derivative on the field variables; for example, a time derivative of $\varphi^{(n)}$ gives,
\begin{align}
    &\frac{d\varphi^{(n)}}{dt} = \sum_{p,q,m} \Bigg(\varphi^{(n),\omega_{p,q,m}}[\tilde{t},x^i] \frac{\partial \phi_j}{\partial t} (-i k^j) e^{-ik^i \phi_i} \notag \\ 
    &+ \Big(\frac{\partial \Omega_j}{\partial t} \frac{\partial \varphi^{(n),\omega_{p,q,m}}}{\partial \Omega_j} + \frac{\partial \delta M_\mu}{\partial t} \frac{\partial \varphi^{(n),\omega_{p,q,m}}}{\partial \delta M_\mu}\Big) e^{-ik^i \phi_i }\Bigg). \label{eq:dphidt}
\end{align}
Evaluating the partial derivatives in Eq.~\eqref{eq:dphidt}: $\frac{\partial \phi_j}{\partial t}=\Omega_j$ using Eq.~\eqref{eq:phi}; $\frac{\partial \Omega_j}{\partial t}=\varepsilon F^{\{0\}}_{\Omega_i}[\Omega_i] +\O(\varepsilon^2)$ from Eq.~\eqref{eq:domega}; and $ \frac{\partial \delta M_\mu}{\partial t}= \varepsilon F_\mu^{\{1\}}[\Omega] + {\cal O}(\varepsilon^2)$ from Eq.~\eqref{eq:dM_mu}. As the background spacetime and background scalar field are stationary, and the perturbations are of the form in Eqs.~\eqref{eq:twotimemetpert1}-\eqref{eq:twotimephi2}, 
we can split all time derivatives into an algebraic $\O(\varepsilon^0)$ piece and a differential $\O(\varepsilon )$ piece,
\begin{align}\label{eq:timederevs}
    \frac{\partial}{\partial t} \rightarrow -ik^j \Omega_j + \varepsilon \frac{\partial}{\partial \tilde t}\ , 
\end{align}
where we have defined the \textit{slow-time derivatives}, $\frac{\partial}{\partial \tilde t}$, such that when it acts on a field variable
\begin{align}
    \frac{\partial}{\partial \tilde t} := F^{\{0\}}_{\Omega_j}[\Omega_i] \frac{\partial \varphi^{(n),\omega_{p,q,m}}}{\partial \Omega_j} + F_A^{\{1\}}[\Omega_j] \frac{\partial \varphi^{(n),\omega_{p,q,m}}}{\partial \delta M_A}.
\end{align}
Note, the slow-time derivatives in Eq.~\eqref{eq:timederevs} contribute at one order in $\varepsilon$  higher than the \textit{fast-time derivatives} (the algebraic part of Eq.~\eqref{eq:timederevs}). We now use a labelling convention~\cite{spiersThesis} to denote the number of slow-time derivatives in a differential operator. Take, for a simple example, the operator
\begin{align}
    A[\varphi^{(n)}]:= \frac{\partial^2 \varphi^{(n)}}{\partial t^2}.
\end{align}
As $\varphi^{(n)}$ can be expressed using Eqs.~\eqref{eq:twotimemetpert1}~and~\eqref{eq:twotimephi2}, we replace the time derivatives with fast and slow-time derivatives using Eq.~\eqref{eq:timederevs}, giving
\begin{align}
    A[\varphi^{(n)}]&= \Big( -ik^j \Omega_j + \varepsilon \frac{\partial}{\partial \tilde t}\Big) \Big( -ik^j \Omega_j + \varepsilon \frac{\partial}{\partial \tilde t}\Big) \varphi^{(n)} \\
    &= -(k^j \Omega_j)^2 \varphi^{(n)} -2i \varepsilon k^j \Omega_j \frac{\partial \varphi^{(n)}}{\partial \tilde t} + \varepsilon^2 \frac{\partial^2 \varphi^{(n)}}{\partial \tilde t^2}.
\end{align}
$A[\varphi^{(n)}]$ can be expressed in orders of slow-time derivatives:
\begin{align}
    A[\varphi^{(n)}]= A^{\langle 0\rangle}&[\varphi^{(n)}] + A^{\langle 1\rangle}[\varphi^{(n)}] + A^{\langle 2\rangle}[\varphi^{(n)}], 
\end{align}
    where
\begin{align}
    A^{\langle 0\rangle}[\varphi^{(n)}] :&=  -(k^j \Omega_j)^2 \varphi^{(n)}, \\
    A^{\langle 1\rangle}[\varphi^{(n)}] :&=  -2i \varepsilon k^j \Omega_j \frac{\partial \varphi^{(n)}}{\partial \tilde t}, \\
    A^{\langle 2\rangle}[\varphi^{(n)}] :&=  \varepsilon^2 \frac{\partial^2 \varphi^{(n)}}{\partial \tilde t^2},
\end{align}
where the number in angular brackets denotes the number of slow-time derivatives in the differential operator.

We also need to expand additional quantities that appear in our equations:
\begin{align}
    z^\alpha &= z_{(0)}^\alpha + z_{(1)}^\alpha + ...\label{eq:zexpansion} \\
    u^\alpha &= u_{(0)}^\alpha + u_{(1)}^\alpha + ... \label{eq:uexpansion}\\
    \Big(\frac{d\tau}{dt}\Big) &= \Big(\frac{d\tau}{dt}\Big)_{(0)} + \Big(\frac{d\tau}{dt}\Big)_{(1)} + ...
\end{align}
with the usual multi-scale expansion definitions~\cite{wardellpoundreview2021}.

The operators in our field equations, Eqs.~\eqref{eq:dGh1},~\eqref{eq:dGh2},~\eqref{eq:first-orderscalarwave}, and~\eqref{eq:second-orderscalarwave}, separate into pieces with various numbers of slow-time derivatives. Take Eq.~\eqref{eq:dGh1}, we can re-express it in the two-timescale approximation as
\begin{align}\label{eq:dGh1TwoTime}
    &\delta G_{ab}^{\langle 0\rangle}[h^{(1)}_{ab}] +  \delta G_{ab}^{\langle 1\rangle}[h^{(1)}_{ab}]\notag \\
    &=8\pi \varepsilon \int_\gamma \mu \Bigg(\Big(u^a_{(0)}u^b_{(0)} + \varepsilon 2 u^{(a}_{(1)\perp}u^{b)}_{(0)}\Big)\frac{\delta^{4}[x^\mu-z_\p^\mu[\tau]]}{\sqrt{-g}} \notag \\
    &- u_a^{(0)}u_b^{(0)}z^\gamma_{(1)\perp} \nabla_\gamma\frac{\delta^{(4)}[x^\mu-z_\p^\mu[\tau]]}{\sqrt{-g}}\Bigg) d\tau_{(0)}  +\order{\varepsilon^3}.
\end{align}
where $z^\gamma_{(1)\perp}=(g^\gamma_{\ \alpha} + u^\gamma_{(0)} u_\alpha^{(0)})z^\alpha_{(1)}$ and $u^\gamma_{(1)\perp}=(g^\gamma_{\ \alpha} + u^\gamma_{(0)}u_\alpha^{(0)})u^\alpha_{(1)}$. Note, Eq.~\eqref{eq:dGh1} is a first order in $\varepsilon$ equation, but implementing the two-timescale approximation has introduced $\varepsilon^2$ pieces in Eq.~\eqref{eq:dGh1TwoTime}. The second order in $\varepsilon$ pieces can be promoted to the second-order field equation, Eq.~\eqref{eq:dGh2}, giving 
\begin{align}\label{eq:dGh1twotime}
    &\delta G_{ab}^{\langle 0\rangle}[h^{(1)}_{ab}]= 8\pi \int_\gamma \mu \frac{\delta^{4}[x^\mu-z_\p^\mu[\tau]]}{\sqrt{-g}}u^a_{(0)}u^b_{(0)} d\tau_{(0)},\\
    &\delta G_{ab}^{\langle 0\rangle}[h^{(2)}_{ab}]= -\delta^2G_{ab}^{\langle 0\rangle}[h^{(1)}_{ab},h^{(1)}_{ab}] - \delta G_{ab}^{\langle 1\rangle}[h^{(1)}_{ab}] \notag \\
    &+\frac{1}{2}\partial^{\langle 0\rangle}_a\varphi^{(1)} \partial^{\langle 0\rangle}_b \varphi^{(1)} -\frac{1}{4}g_{ab}(\partial^{\langle 0\rangle}\varphi^{(1)})^2 \notag \\
     &+ 4\pi \int_\gamma \frac{\delta^{4}[x^\mu-z_\p^\mu[\tau]]}{\sqrt{-g}}\Big(2m_{[1]}\varphi^{(1)}_R u^{(0)}_au^{(0)}_b +\notag \\
    &\mu \big(4h^{R(1)}_{ac}{(0)}^c{(0)}_b-{(0)}_a {(0)}_b (g^{cd}_{(0)} - u_{(0)}^{c}u_{(0)}^{d})h^{R(1)}_{cd}   \big) \Big) d\tau_{(0)} \notag \\
    &+16\pi\int_\gamma \mu \bigg( 2u^{(0)\perp}_{(a}u^{(0)}_{b)} \frac{\delta^{(4)}[x^\mu-z_\p^\mu[\tau]]}{\sqrt{-g}} \notag \\
    &- u_a^{(0)}u_b^{(0)}z^\gamma_{(1)\perp} \nabla_\gamma\frac{\delta^{(4)}[x^\mu-z_\p^\mu[\tau]]}{\sqrt{-g}} \bigg)d\tau_{(0)}    ,\label{eq:dGh2TT}
\end{align}
where we have also expanded $\delta^2 G_{ab}[h^{(1)}_{ab}]$ and $\partial_\mu$ in orders of slow-time derivatives.

As we are interested in first-post adiabatic modelling here, we have neglected any terms that are $\O(\varepsilon^3)$. For example, $\delta^2 G^{\langle 1\rangle}_{ab}[h^{(1)}_{ab}]=\O(\varepsilon^3)$; generally $\delta^n G^{\langle m\rangle}_{ab}[h^{(i)}_{ab}]=\O(\varepsilon^{(ni+m)})$. Deriving higher-order equations with this algorithm is trivial.

Applying the slow-time derivative expansion algorithm to the scalar perturbation field equations, Eqs.~\eqref{eq:first-orderscalarwave}~and~\eqref{eq:second-orderscalarwave}, give
\begin{align}\label{eq:first-orderscalarwaveTwotime}
    \Box&^{\langle 0\rangle} \varphi^{(1)} = 16\pi \int_\gamma m_{[1]}\frac{\delta^{(4)}[x^\mu-z_\p^\mu[\tau]]}{\sqrt{-g}}d\tau_{(0)}, \\
    \Box&^{\langle 0\rangle} \varphi^{(2)} =-\frac{8\pi\alpha^{(n)}}{\sqrt{-g}}\G^{(0)} -\Box^{\langle 1\rangle} \varphi^{(1)} \notag \\
    &- h^{ab}_{(1)} \nabla_a\nabla_b\varphi^{(1)} -(\nabla^a h^{(1)}_{ab})\nabla^b\varphi^{(1)} +\frac{1}{2} (\nabla^b h_{(1)}) \nabla_b \varphi^{(1)}    \notag \\
    &+16\pi \Bigg( \int_\gamma m_{[2]}\varphi^{(1)}_R\frac{\delta^{(4)}[x^\mu-z_\p^\mu[\tau]]}{\sqrt{-g}}d\tau_{(0)} \notag \\ &-\frac{1}{2} \int_\gamma m_{[1]}\frac{\delta^{(4)}[x^\mu-z_\p^\mu[\tau]]}{\sqrt{-g}} (g^{ab}+u_{(0)}^au_{(0)}^b)h_{ab}^{R(1)}d\tau_{(0)}  \notag\\
    &-\int_\gamma m_{[1]}z^\gamma_{(1)\perp} \nabla_\gamma\frac{\delta^{(4)}[x^\mu-z_\p^\mu[\tau]]}{\sqrt{-g}} d\tau_{(0)}
    \Bigg)\ .\label{eq:second-orderscalarwaveTwotime}
\end{align}

From the field variable coefficients, $\varphi^{(1),\omega_{p,q,m}}$, $\varphi^{(2),\omega_{p,q,m}}$, $h^{(1),\omega_{p,q,m}}_{ab}$, and $h^{(2),\omega_{p,q,m}}_{ab}$, one can calculate the orbit averaged self-force: 
\begin{align}\label{eq:STSF}
F^a_{SF}&=\varepsilon F_{(1)}^a[h^{(1),\omega_{p,q,m}}_{ab},\varphi^{(1),\omega_{p,q,m}}]\notag \\ & \ \ \ + \varepsilon^2 F_{(2)}^a[h^{(2),\omega_{p,q,m}}_{ab},\varphi^{(2),\omega_{p,q,m}}] +\order{\varepsilon^3}.
\end{align}
Where we have extended Eq.~\eqref{eq:SF} to include the self-force from the scalar field. We use the arguments in Ref.~\cite{hinderer2008} to split the self-force into dissipative and conservative pieces in a post-adiabatic expansion,
\begin{align}\label{eq:STSFPA}
F^a_{SF}&=F_{\{0\}(1)\diss}^a[h^{(1),\omega_{p,q,m}}_{ab},\varphi^{(1),\omega_{p,q,m}}]\notag \\ 
& \ \ \ +  F_{\{1\}(1)\cons}^a[h^{(1),\omega_{p,q,m}}_{ab},\varphi^{(1),\omega_{p,q,m}}] \notag \\ 
& \ \ \ + F_{\{1\}(2)\diss}^a[h^{(2),\omega_{p,q,m}}_{ab},\varphi^{(2),\omega_{p,q,m}}]+  ...
\end{align}
where the numbers in the curly brackets denotes the post-adiabatic order of the self-force contribution.

We can input Eq.~\eqref{eq:STSFPA} into Eq.~\eqref{eq:phiF} to calculate the phase evolution. In practice, the phases evolve via Eq.~\eqref{eq:phi}; that is, the evolution of the orbital frequencies is determined by the self-force coefficients $F_{\Omega_i}^{\{0\}}$ and $F_{\Omega_i}^{\{1\}}$ in Eq.~\eqref{eq:domega}. The self-force coefficients $F_{\Omega_i}^{\{n\}}$ and $F_A^{\{1\}}$ can be determined from the field variable coefficients (similarly to Eq.~\eqref{eq:STSFPA}),
\begin{align}\label{eq:FOmega0}
    F_{\Omega_i}^{\{0\}}[\Omega_j] &= F_{\Omega_i}^{\{0\}}[h^{(1),\omega_{p,q,m}}_{ab},\varphi^{(1),\omega_{p,q,m}}], \\
    F_{\Omega_i}^{\{1\}}[\Omega_j,\delta M_A] &=  F_{\Omega_i}^{\{1\}}[h^{(1),\omega_{p,q,m}}_{ab},\varphi^{(1),\omega_{p,q,m}},\notag \\ & \ \ \ \ \  \ \ \ \ \  \ \ \ \ \ \ \ \ h^{(2),\omega_{p,q,m}}_{ab},\varphi^{(2),\omega_{p,q,m}}],\label{eq:FOmega1}\\
    F_A^{\{1\}}[\Omega_j]&=F_A^{\{1\}}[h^{(1),\omega_{p,q,m}}_{ab},\varphi^{(1),\omega_{p,q,m}}].\label{eq:FA0}
\end{align}

The force coefficients in Eqs.~\eqref{eq:FOmega0} and~\eqref{eq:FOmega1} can be derived from the perturbative equations of motion, Eq.~\eqref{eq:a1}~and~\eqref{eq:a2}. However, one must account for slow-time derivative contributions to Eq.~\eqref{eq:a2}. To find the additional contribution one can examine Eq.~\eqref{eq:a1} using Eqs.~\eqref{eq:timederevs},~\eqref{eq:zexpansion}, and~\eqref{eq:uexpansion}, giving
\begin{align}
    &a_{(2)\mathrm{slow}}^a = m_{[1]}^{(1)} (g^{ab} + u_{(0)}^a u_{(0)}^b +2u^{(a}_{(1)}u^{b)}_{(0)})\partial^{\langle 1\rangle}_b \varphi^{(1)}_{\mathcal{R}}\notag \\ 
    &   -\frac{1}{2} (g^{ab}+u_{(0)}^au_{(0)}^b)\big(2\partial^{\langle 1\rangle}_e h^{(1)\mathcal{R}}_{bd} -\partial^{\langle 1\rangle}_b h^{(1)\mathcal{R}}_{be}\big)u_{(0)}^du_{(0)}^e \notag \\
    &-\frac{1}{2} (g^{ab} +u_{(0)}^au_{(0)}^b)(2h^{(1)\R}_{bd;e}-h^{(1)\R}_{de;b})2u^{(d}_{(1)} u^{e)}_{(1)}.\label{eq:a2slow}
\end{align}
The second-order equation of motion then becomes
\begin{align}
    a^a_{(2)}= a^{a}_{(2)\grav}+a^{a}_{(2)\scal}+a_{(2)\mathrm{slow}}^a.\label{eq:a2full}
\end{align}
From Eqs.~\eqref{eq:a1} and~\eqref{eq:a2full} one can derive precise relations for Eqs.~\eqref{eq:FOmega0} and~\eqref{eq:FOmega1}. The force coefficients in Eq.~\eqref{eq:FA0} can be determined from the orbit averaged fluxes entering the primary black hole horizon in terms of $h^{(1),\omega_{p,q,m}}_{ab}$, and $\varphi^{(1),\omega_{p,q,m}}$. 

This completes the two-timescale expansion of our first-post adiabatic accuracy modelling scheme for non-spinning binaries in scalar-tensor theories of gravity.

\section{Summary and Conclusions}

In this paper, we have derived a framework for computing the premiere first-post adiabatic binary modelling scheme in theories with a massless scalar field non-minimally coupled to gravity. As discussed in detail in Sec.~\ref{sec:pointcharge}, our main assumptions are that any interaction terms for the scalar are suppressed by a coupling with mass-dimension-2 or higher, the scalar field is shift symmetric, and the solutions of the theory continuously connect to those of general relativity with a minimally coupled scalar field.
We have used perturbative scaling arguments, using the suppression of black hole scalar charge by its mass, to isolate the adiabatic and first-post adiabatic contributions. This allows us to ignore most of the general coupling between the metric and scalar field. This approach is similar to Refs.~\cite{maselli2020detecting, Maselli:2021men, Barsanti:2022ana} who developed and implemented this method to adiabatic order.

We have produced an ansatz for the matter action of a compact object with a scalar charge, which is the foundation of our formalism. The long-established scalar charged point particle action, Eq.~\eqref{eq:skeleton1}~\cite{damour1992tensor}, becomes problematic in the self-force approach because of divergences in the field perturbations on the worldline. We have combined the scalar charged point particle action with the effective metric approach~\cite{detweiler2012second-order} to produce our ansatz for the particle action, Eq.~\eqref{eq:effectiveskeleton}. The effective metric and effective scalar field are regular on the worldline, making our action well defined. In the limit $\varphi\rightarrow0$, Eq.~\eqref{eq:effectiveskeleton} is equivalent to a point-mass effective action in GR, which is consistent with the first- and second-order self-force method~\cite{misata1997, quwa1997, pound2012second, gralla2012, Upton, detweiler2012second-order}. Our ansatz and formalism could be checked by calculating a matched asymptotic expansion~\cite{kevorkian2012multiple, eckhaus2011asymptotic, kates1981underlying, pound2010singular, pound2012second} for a scalar charged compact binary in scalar-tensor theories of gravity. 

In Sec.~\ref{sec:perteqs}, we derived field equations for the first- and second-order metric and scalar perturbations, Eqs.~\eqref{eq:dGh1},~\eqref{eq:dGh2},~\eqref{eq:first-orderscalarwave}, and~\eqref{eq:second-orderscalarwave}. We also derive the first- and second-order equations of motion for the secondary object, Eq.~\eqref{eq:a1}~and~\eqref{eq:a2}. Our formalism also trivially extends to higher orders. Because Eq.~\eqref{eq:dGh1} is identical in GR, the first-order Teukolsky equation holds in our formalism~\cite{teuk1972, teuk1973}. It is straightforward to convert the second-order metric perturbation field equations (Eqs. \eqref{eq:dGh1twotime} or~\eqref{eq:dGh2TT}) and convert them to second-order Teukolsky equations by applying the methods used in Refs.~\cite{mySecond-orderTeuk, green2020teukolsky, hussain2022approach}. Ref.~\cite{li2023perturbations} presents an alternative method for deriving a Teukolsky equation in alternative theories of gravity. Also, the metric reconstruction methods developed in GR hold in our formalism (see CCK~\cite{chrzanowski1975vector, cohen1975space}, GHZ~\cite{green2020teukolsky, toomani2021new}, AAB~\cite{aksteiner2019new}, and Lorenz gauge~\cite{dolan2022gravitational} metric reconstruction).

In Sec.~\ref{sec:two-timescale}, we integrate our formalism into the two-timescale approximation, allowing for efficient first-post adiabatic calculations. Our formalism is also consistent in the time-domain~\cite{miller2020two} and the self-consistent approximation~\cite{pound2015}. An additional modelling problem that needs addressing is resonances~\cite{van2014conditions}. We expect that the methods used to address resonances in GR, such as those implemented in Ref.~\cite{nasipak2021resonant}, will also be applicable to our formalism.

Remarkably, our formalism has added only one additional parameter, $m_{[2]}^{(1)}$, to the adiabatic order formalism of Refs.~\cite{maselli2020detecting}. Hence, it appears that the scalar charge $d$ and  $m_{[2]}^{(1)}$ capture the effects of the scalar field to first post-adiabatic order in a very large class of theories. Our understanding of $m_{[2]}^{(1)}$ is currently limited, and it would be interesting to investigate how different theories of gravity generate a non-zero $m_{[2]}^{(1)}$ and how significant this contribution is to an EMRI model. It is conceivable that in a subset of theories, the only significant extra parameter to the first post-adiabatic order is indeed  $d$. It would also be interesting to extend our formalism to include a mass of the scalar field and interactions that do not respect shift symmetry. The effect of the former has already been studied at adiabatic order in \cite{Barsanti:2022ana}. We also intend to publish a follow-up paper which will extend our formalism to include the first-post adiabatic effects of the spin and scalar dipole of the secondary compact object.


The main motivation for this work is to model EMRI waveforms to first-post adiabatic accuracy for LISA data analysis. 
Accuracy requirement arguments suggest that calculating the second-order self-force to $\sim 1\%$ accuracy will likely be sufficient for LISA data analysis~\cite{Miller:2020bft}. If the second-order scalar self-force (and the effect of the scalar field on the gravitational second-order self-force) are suppressed by two orders of magnitude compared to the gravitational self-force, then their effect may be neglected. Ref.~\cite{maselli2020detecting} found the adiabatic scalar self-force is $\O( 1\%)$ of the gravitational self-force for $d=0.3$; for smaller $d$ the scalar self-force is further suppressed. If a similar relationship is found at first-post adiabatic order, then the conservative piece of the first-order scalar self-force and the dissipative piece of the second-order scalar self-force may be negligible. Nevertheless, the adiabatic contribution (the dissipative piece of the first-order self-force) will still be significant. That is, waveforms in scalar-tensor theories of gravity will be significantly different from those in GR, but it may be even easier to model binaries in scalar-tensor theories of gravity than our formalism suggests.



Implementing our formalism will be a similar difficulty as computing first-post adiabatic models in GR. The most challenging part of such calculations is calculating $\delta^2G[h^{(1)}_{ab},h^{(1)}_{ab}]$ near the worldline. We have shown that this calculation is identical in GR and our formalism. A method for calculating $\delta^2G[h^{(1)}_{ab},h^{(1)}_{ab}]$ near the worldline is given in Ref.~\cite{miller2016second}, but it is inefficient and highly technical. There are additional, similarly challenging to calculate, pieces in our formalism (again containing products of divergences on the worldline): $\frac{1}{2}\partial_a\varphi^{(1)} \partial_b \varphi^{(1)} -\frac{1}{4}g_{ab}(\partial\varphi^{(1)})^2$ in Eq.~\eqref{eq:dGh2} and $h^{ab}_{(1)} \nabla_a\nabla_b\varphi^{(1)}-(\nabla^a h^{(1)}_{ab})\nabla^b\varphi^{(1)} +\frac{1}{2} (\nabla^b h_{(1)}) \nabla_b \varphi^{(1)}$ in Eq.~\eqref{eq:second-orderscalarwave}. We propose again using the method in Ref.~\cite{miller2016second}, or any new methods developed to tackle the problem in GR (this is currently an active area of research in the self-force community).

Our formalism could also be used to produce intermediate-mass-ratio inspiral waveforms. The results in Refs.~\cite{pound2020second, wardell2021gravitational, warburton2021gravitational} show encouraging agreement with Numerical Relativity and first-post adiabatic self-force waveforms in GR for quasi-circular inspirals of Schwarzschild black holes in the mass-ratio regime of $1:10$. These results will soon be used to help future LVK data analysis (as LVK begins to probe deeper into the disparate mass ratio regime). Implementing our formalism, even for the simpler case of non-spinning (or linear in spin) black hole primary, would be an important step towards detecting or constraining the existence of a new fundamental scalar field. 

\appendix

\section{Self-force formalism in GR}\label{app:SFGR}

In this appendix, we summarise the self-force approach 
for binary black holes in GR. We focus on the relevance of 
first-post adiabatic waveform models for gravitational wave 
observations. Additionally, we show how the first- and second-order field equations and equations of motion can be derived from an effective action.

In black hole perturbation theory, the metric ($\g_{ab}$) 
is expressed as an expansion in orders of a small 
parameter, as in Eq.~\eqref{eq:gExpansion}. In the perturbative self-force approach, the small parameter is the mass ratio, $\varepsilon=\mu/M$.
Working in GR, $g^{(0)}_{ab}$ is a solution to the Einstein 
field's equations. Here, we take $g^{(0)}_{ab}$ to be the Kerr metric~\cite{kerr1963gravitational}.


The presence of the secondary compact object produces the metric perturbations. These perturbations impart a force on the compact object, causing it to move off geodesic motion in the background spacetime. The so-called self-force per unit mass ($F^a_{SF}$) can also be expressed as an expansion in orders of the mass ratio: 
\begin{align}\label{eq:SF}
F^a_\textnormal{SF}=\varepsilon F_{(1)}^a\big[h^{(1)}_{ab}\big] + \varepsilon^2 F_{(2)}^a\big[h^{(2)}_{ab}\big] +\order{\varepsilon^3}.
\end{align}
\noindent The effect of the self-force on the motion of the inspiraling object is described by the equation of motion,
\begin{align}\label{eq:EOM}
u^b \nabla_b u^a=a^a=(\varepsilon F_{(1)}^a + \varepsilon^2 F_{(2)}^a) +\order{\varepsilon^3},
\end{align}
where $u^a$ is the four-velocity of the compact object in the background spacetime, $a^a$ is the four-acceleration, and $\nabla_a$ is the covariant derivative of the background metric. Eq.~\eqref{eq:EOM} is given by the MiSaTaQuWa equation~\cite{misata1997, quwa1997} to first order, and by the second-order equation of motion~\cite{pound2012second, gralla2012} to second order. 

\subsection{The self-force action approach}

\label{sec:AA1}

Here, we show that it is possible to derive the GR self-force equations of motion and field equations, up to second-order, directly from an action. We begin with the action,
\begin{align}\label{eq:GRaction}
    S[\g_{ab} ,\Psi] = S_{EH}[\g_{ab} ] +  S_\m[\g_{ab} , \Psi]\ ,
\end{align}
where $S_\m$ is the matter action, $\Psi$ are the matter fields, and 
\begin{align}\label{eq:SEH}
    S_{EH}[\g_{ab} , \varphi]=\int \frac{\sqrt{-\g}}{16\pi} R  \ d^4x\ ,
\end{align}
is the Einstein--Hilbert action. 

In the compact binary problem, spacetime is a vacuum everywhere except on the position of the compact object (the worldline). The matter action in Eq.~\eqref{eq:GRaction} can be replaced by the effective point particle action
\begin{align}\label{eq:effectiveskeletonGR}
    S_\p=-\int_{\gamma} \mu d\tilde{s} = - \int_{\gamma} \mu \sqrt{\tilde{g}_{ab} \tilde{u}^a\tilde{u}^b}d\tilde{\tau}.
\end{align}
where $\tilde g_{ab}$ is the effective metric defined in Eq.~\eqref{eq:tildephiexpansion}, $\tilde\tau$ is the proper time in the effective 
spacetime, and $\tilde{u}^\alpha=dz^\alpha/d\tilde{\tau}$.

Varying the action~\eqref{eq:GRaction} with respect to the 
body's path, $x^\mu\rightarrow x^\mu +\delta x^\mu$, results in the equation of motion
\begin{align}\label{eq:effectiveaGR}
   \mu \tilde{a}^a=0 ,
\end{align}
where $\tilde{a}^a=\tilde{u}^b\tilde{\nabla}_b\tilde{u}^a$.

Expanding Eq.~\eqref{eq:effectiveaGR} using the expansions Eq.~\eqref{eq:tildephiexpansion}, and those in App.~\ref{app:expansiontools}, one recovers the MiSaTaQuWa equation,
\begin{equation}\label{eq:a1gravGR}
     a^a_{(1)} =  -\frac{1}{2} (g^{a}_{(0) c}+u^au_c)(2h^{(1)\R}_{d(b;e)}-h^{(1)\R}_{be;d})u^b u^e\ ,
\end{equation}
and the second-order equation of motion,
\begin{align}\label{eq:a2gravGR}
    & a^a_{(2)} = -\frac{1}{2} \Big[(g^{a}_{(0) c}+u^au_c)(2h^{(2)\mathcal{R}}_{d(b;e)}-h^{(2)\mathcal{R}}_{be;d})  \notag \\ & \ \ \ \ \ + (g^{a}_{(0)c}+u^au_c)h_{(1)\mathcal{R}}^{ cd}(2h^{(1)\mathcal{R}}_{d(b;e)}-h^{(1)\mathcal{R}}_{be;d}) \Big] u^b u^e \ .
\end{align}

Additionally, the field equations can be derived by varying the action~\eqref{eq:GRaction}. This process is more delicate because one must vary the $S_{EH}[\g_{ab} ]$ with respect to the metric ($\g_{ab}$) and $S_\m[\g_{ab} ]=S_p[\tilde g_{ab}]$ with respect to $\tilde g_{ab}$. The resulting field equation is~\cite{detweiler2012second-order, Upton}
\begin{align}\label{eq:GRfieldeq}
    G[\g_{ab}]=8\pi \int_\gamma \mu\frac{\delta^{4}[x^\mu-z_\p^\mu[\tilde\tau]]}{\sqrt{-\tilde{g}}}\tilde{u}^a\tilde{u}^b d\tilde{\tau}.
\end{align}
Geroch's theorem~\cite{geroch1970domain} tells us Eq.~\eqref{eq:GRfieldeq} is not formally well defined as a partial differential equation, but it will be useful for deriving the correct perturbative equations (to at least second-order). A perturbative expansion of Eq.~\eqref{eq:GRfieldeq} using the expansions in Eq.~\eqref{eq:gExpansion},~\eqref{eq:tildephiexpansion}, and the those in App.~\ref{app:expansiontools}, gives the first- and second order field equations: 
\begin{align}\label{eq:dGh1GR}
    \delta G_{ab}[h^{(1)}_{ab}]&= 8\pi \int_\gamma \mu \frac{\delta^{4}[x^\mu-z_\p^\mu[\tau]]}{\sqrt{-g}}u^au^b d\tau , \\
   \delta G_{ab}[h^{(2)}_{ab}] &= -\delta^2G_{ab}[h^{(1)}_{ab},h^{(1)}_{ab}] \notag \\
    &   + 4\pi \int_\gamma \frac{\delta^{4}[x^\mu-z_\p^\mu[\tau]]}{\sqrt{-g}}\notag \\
    &\mu \bigg(   4h^{\R(1)}_{ac}u^cu_b- u_a u_b (g^{cd}_{(0)} - u^{c}u^{d})h^{R(1)}_{cd} \bigg) d\tau .\label{eq:dGh2GR}
\end{align}

\subsection{Why second-order self-force and first-post adiabatic accuracy}

Next, we summarise which parts of the self-force expansion are required for an accurate model. Summarising Ref.~\cite{hinderer2008}, we characterise an accurate model as having a small error on the final position of the compact object over the course of the inspiral. An inspiral is generally considered to evolve on a so-called \textit{slow-timescale} ($\tilde t$) related to the radiation reaction timescale, $\tilde t\sim t_{rr}\sim \frac{M}{\varepsilon }=\O(\frac{1}{\varepsilon })$~\cite{miller2020two}. This relation can be derived by considering the rate at which orbital energy is dissipated from the system through gravitational waves. The rate of energy dissipation scales as $\dot{E}=\frac{dE}{dt}\sim (h^{(1)}_{ab})^2\sim \varepsilon^2$ \cite{andersson2019gravitational}. The orbital energy of the compact object is $E\sim \mu$. As $t\sim\frac{E}{\dot{E}}$ and $\frac{E}{\dot{E}}\sim \frac{\mu}{\varepsilon^2}\sim \frac{M}{\varepsilon}$; therefore, $t\sim \frac{M}{\varepsilon}$. 

The error in the final position ($\delta z^\mu$) relates to the slow-timescale and the error in the acceleration ($\delta a^\mu$)~\cite{pound2012field}:
\begin{align}
\delta z^\mu\sim t^2 \delta a^\mu \sim \tfrac{1}{\varepsilon^2} \delta a^\mu.
\end{align}
\noindent For $\delta z^\mu$ to be small, $\delta a^\mu = \O( \varepsilon^3)$ is necessary. Hence, Eq.~\eqref{eq:EOM} must be calculated through second order to achieve an accurate model over an entire inspiral. That is, the first- and second-order self-force is required. 

In practice, the position of the compact object is expressed using orbital phases, $\phi_i$. For generic orbits in Kerr, there are three orbital phases, that is, $i=\{1,2,3\}$, or $i=\{r,\theta,\phi\}$. The phases obey the expansion~\cite{hinderer2008}
\begin{align}\label{eq:phi-expansion}
    \phi_i[t, \varepsilon]=\frac{1}{\varepsilon}\phi^{\{0\}}_i[t, \varepsilon] + \phi^{\{1\}}_i[t, \varepsilon] +\mathcal{O}(\varepsilon).
\end{align}
For the error in Eq.~\eqref{eq:phi-expansion} to be small as $\varepsilon\rightarrow 0$ the coefficients $\phi^{\{1\}}_i[t, \varepsilon]$ and $\phi^{\{0\}}_i[t, \varepsilon]$ must be known. The appearance of $\frac{1}{\varepsilon}$ terms in the expansion of $\phi_i[t, \varepsilon]$ is acquired from the inspiral evolving over a slow-timescale. 

Ref.~\cite{hinderer2008} showed that $\phi^{\{0\}}_i[t, \varepsilon]$, the so-called \textit{adiabatic contribution}, depends on the dissipative piece of the first-order self-force. They additionally showed that $\phi^{\{1\}}_i[t, \varepsilon]$ depends on the conservative piece of the first-order self-force and the dissipative piece of the second-order self-force:
\begin{align}\label{eq:phiF}
    \phi^{\{0\}}_i[t, \varepsilon]&=\phi^{\{0\}}_i\big[F_{(1)\diss}^a[h^{(1)}_{ab}]\big], \\
    \phi^{\{1\}}_i[t, \varepsilon]&=\phi^{\{1\}}_i\big[F_{(1)\cons}^a[h^{(1)}_{ab}],F_{(2)\diss}^a[h^{(2)}_{ab}]\big].\label{eq:phiF1}
\end{align}
The reasoning for the conservative self-force being suppressed by one order in $\varepsilon$ is the conservative self-force averages out over a generic Kerr geodesic~\cite{hinderer2008}. 
Eqs.~\eqref{eq:phi-expansion},~\eqref{eq:phiF}, and~\eqref{eq:phiF1} show first-post adiabatic accurate models require the full first-order self-force and the dissipative piece of the second-order self-force.

Ref.~\cite{hinderer2008} also provides the framework for implementing a two-timescale approximation to produce first-post adiabatic 
self-force binary waveforms models in GR.
Recently, first-post adiabatic waveforms have shown incredible 
agreement with Numerical Relativity waveforms  
for quasi-circular inspirals of Schwarzschild black holes, 
even in the $1:10$ mass ratio regime~\cite{pound2020second, wardell2021gravitational, warburton2021gravitational}. These 
ground-breaking results suggest that first-post adiabatic models 
will play a key role in future gravitational wave science, 
across a mass-ratio range much wider than expected. 


\section{Necessary expansions}\label{app:expansiontools}

In Eq.~\eqref{eq:effectiveTp}, $\tilde{g}$ appears explicitly in $\sqrt{-\tilde{g}}$ and implicitly in $\tilde{u}^a$ and $d\tilde{\tau}$. We expand each $\tilde{g}$ dependence perturbatively around $g_{ab}^{(0)}$ as follows~\cite{Upton},
\begin{align}
    \frac{1}{\sqrt{-\tilde{g}}}&=\frac{1}{\sqrt{-g}}(1-\frac{\varepsilon }{2}g^{ab}h^{R(1)}_{ab})+\order{\varepsilon^2},  \label{eq:sqrtdetgexpansion} \\    
    \frac{d\tau}{d\tilde{\tau}}&=\frac{1}{\sqrt{1-h^R_{ab}u^au^b}}\notag \\&=1+\frac{\varepsilon }{2}h^{R(1)}_{ab}u^au^b  +\frac{3\varepsilon^2}{8}[h^{R(1)}_{ab}u^au^b]^2+\order{\varepsilon^3},\label{eq:dtautildeexpanaion}\\    
    \frac{d\tilde\tau}{d\tau}&=\sqrt{1-h^R_{ab}u^au^b} =1-\frac{\varepsilon }{2}h^{R(1)}_{ab}u^au^b +\order{\varepsilon^2},\label{eq:dtildetauexpanaion}
\end{align}
noting\footnote{We require the expansion in Eq.~\eqref{eq:dtautildeexpanaion} to an order higher than Eqs.~\eqref{eq:sqrtdetgexpansion} and~\eqref{eq:dtildetauexpanaion} for the expansion of Eq.~\eqref{eq:effectivea}.}, $\tilde{u}^a=\frac{d\tau}{d\tilde{\tau}}u^a$ and $\tilde{\tau}=\frac{d\tilde\tau}{d\tau}d\tau$.

$T_{ab}^{\m }$ appears in Eq.~\eqref{eq:EFEST} with indices down whereas in Eq.~\eqref{eq:effectiveTp} it is expressed with indices up. Ref.~\cite{Upton} showed that the indices of the stress-energy tensor are raised and lowered by the effective metric (not the background metric). That is,
\begin{align}\label{eqTdownab}
   T^\m_{ab}&=\tilde{g}_{ac}\tilde{g}_{bd}T_\m^{ab} =\varepsilon g^{(0)}_{ac}g^{(0)}_{bd}T_{(1)}^{ab}\notag\\
    & + \varepsilon^2\left[ g^{(0)}_{ac}g^{(0)}_{bd}T_{\m(2)}^{ab} +2h^{R(1)}_{a(c} g_{d)b}^{(0)} T^{ab}_{\m(1)}\right] + \order{\varepsilon^3} .
\end{align}

\begin{acknowledgments}
AS and TS acknowledge the partial support
from the STFC Consolidated Grant no. ST/V005596/1.
AS would like to thank Adam Pound for helpful discussions and comments.
\end{acknowledgments}

\bibliography{bib.bib}
\bibliographystyle{apsrev4-1}
\end{document}